\documentclass[10pt]{iopart}

\usepackage{iopams}

\expandafter\let\csname equation*\endcsname\relax 
\expandafter\let\csname endequation*\endcsname\relax 

\usepackage{amsmath}

\usepackage[font={footnotesize}]{caption}

\usepackage{graphicx}
\usepackage{dcolumn}
\usepackage{bm}
\usepackage{xcolor}

\begin{document}

\title{Probing  the spectral dimension of quantum network geometries}

\author{Johannes Nokkala}
\address{Instituto de F\'{i}sica Interdisciplinar y Sistemas Complejos (IFISC, UIB-CSIC), Campus Universitat de les Illes Balears E-07122, Palma de Mallorca, Spain}
\ead{johannes@ifisc.uib-csic.es}

\author{Jyrki Piilo}
\address{Turku Centre for Quantum Physics, Department of Physics and Astronomy,
University of Turku, FI-20014 Turun yliopisto, Finland}
\ead{jyrki.piilo@utu.fi}

\author{Ginestra Bianconi}
\address{School of Mathematical Sciences, Queen Mary University of London, London, E1 4NS, United Kingdom\\
Alan Turing Institute, The British Library, London, United Kingdom}
\ead{ginestra.bianconi@gmail.com}

\date{\today}

\begin{abstract}
We consider an environment for an open quantum system described by a ``Quantum Network Geometry with Flavor" (QNGF) in which the nodes are coupled quantum oscillators. The  geometrical nature  of QNGF is reflected in the  spectral properties  of the  Laplacian matrix of the  network which display a finite spectral dimension, determining also the frequencies of the  normal modes of QNGFs. We show that an \textit{a priori} unknown spectral dimension can be indirectly estimated by coupling an auxiliary open quantum system to the network and probing the normal mode frequencies in the low frequency regime. We find that the network parameters do not affect the estimate; in this sense it is a property of the network geometry, rather than the values of, e.g., oscillator bare frequencies or the constant coupling strength. Numerical evidence suggests that the estimate is also robust both to small changes in the high frequency cutoff and noisy or missing normal mode frequencies. We propose to couple the auxiliary system to a subset of network nodes with random coupling strengths to reveal and resolve a sufficiently large subset of normal mode frequencies.
\end{abstract}

%
%

\section{\label{sec:introduction}Introduction}

Networks \cite{Laszlo,Newman2010} describe discrete topologies that can capture the architecture of complex systems, from the brain to societies. As an increasing number of  datasets about complex systems are becoming available, the characterization of real-world network structures has significantly enriched the understanding about the relation between network topology and dynamics. A classic result of Network Theory is that  the statistical properties of the complex network topology, including for instance the degree distribution, strongly  affect the properties of dynamical processes such as  percolation, epidemic spreading and spin models \cite{Doro_crit}. However only recently the scientists have realized that a large variety of network datasets have an intrinsic geometrical nature that affects their dynamical properties.

The term ``network geometry" \cite{Bianconi_2015,Boguna} refers to discrete topological spaces with notable geometrical features. These structures can be modelled by using simplicial complexes \cite{Bianconi_2015,Lambiotte} which are discrete structures not only formed by nodes and links but formed also by triangles, tetrahedra and so on. A general feature of network geometries is that they display a finite spectral dimension \cite{Toulouse,Burioni1}, i.e. the random walks defined on these structures relax to its stationary state only algebraically like in finite dimensional Euclidean lattices. The spectral dimension of an $d$-dimensional Euclidean lattice is $d$, however it is well known that also fractals have a finite spectral dimension \cite{Toulouse},  and only recently it  has been shown that several real network datasets also display this important geometrical property \cite{Emergent}.

Recently a general theoretical framework called ``Network Geometry with Flavor" (NGFs) \cite{NGF,bianconi2017emergent} has been proposed to model network geometries using simplicial complexes. The NGFs depend on the dimension $d$ of their building blocks and on another parameter $s$ called {\em flavor}. For any dimension $d$ and any flavor $s$ the NGFs capture the main characteristics of complex networks including  modularity, small-world properties and heterogeneous degree distribution. Moreover these topologies display also a finite spectral dimension which is the signature of their geometrical nature. It is also worth noting that the spectral dimension can change significantly the  dynamical properties of classical and quantum random walks \cite{Burioni1,childs2004spatial,Boettcher}, and the Kuramoto model leading to an anomalous frustrated synchronization \cite{Ana1,Ana2}.

During the recent years, Network Theory has also opened significant possibilities within the framework of quantum physics~\cite{Bianconi_2015,Biamonte_2019}. Indeed, the concept and use of \textit{quantum complex networks} are becoming increasingly common. Here, the interest is focused, e.g., on quantum correlations and phase transitions~\cite{PhysRevLett.119.225301}, quantum communication networks and internet~\cite{Brito_2020}, quantum walks~\cite{PhysRevX.3.041007}, and generalizing the classical concepts of Network Theory to quantum domain~\cite{Perseguers_2010,PhysRevX.4.041012}.

Our current interest lies on using quantum complex networks within the framework of open quantum systems, where the environment---that an open system interacts with---consists of a network of interacting quantum entities.
Intriguing possibilities are now provided by using an open system to probe the properties of the quantum complex network. 
Here, one assumes that the only available information about the quantum network is the information that can be inferred by  studying  the properties of the probe system only.  This approach has been fully developed  in the case in which the quantum network is formed by a network of  harmonic quantum oscillators coupled to an open system oscillator that acts as the probe~\cite{nokkala2016complex,nokkala2018local}.  Moreover, the experimental  implementation of the theoretical framework and the probing schemes, have been recently proposed using a multi-mode optical set-up~\cite{nokkala2018reconfigurable}.

Building on these results, in this paper we investigate the QNGF which is the environment of a quantum open system (the probe) and is formed by a network of quantum harmonic oscillators coupled according to the topology of the NGF.

Quantum algorithms to investigate the topological properties of simplicial complexes have been proposed in previous works \cite{Garnerone} however not using the theory of open quantum systems.

We are interested in how to obtain crucial information about the geometrical properties of the QNGF without having any \textit{a priori} knowledge of the generating mechanism of the NGF or of its structure.  In particular our main goal is to infer the value of the spectral dimension of the underlying topology of the QNGF.
 We provide a connection between the eigenfrequencies of the QNGF  and the eigenvalues of the Laplacian matrix of  the NGF. Using the theory of open quantum systems combined with data science  we are able to probe  the value of the spectral dimension of the QNGF. Moreover, we also consider cases where the probe has limited capacity to reveal the full set of eigenfrequencies of the oscillator network, and also when the obtained data is noisy, i.e., influenced by random fluctuations of the eigenfrequencies. The results indeed demonstrate that the probing scheme for the spectral dimension is rather robust when not having fully ideal setting at hand.

The paper is organized as follows. We introduce the considered quantum network model in Sections~\ref{sec:NGF} and~\ref{sec:QNGFsub}, focusing on the topology and physical model, respectively, { and introduce the open quantum system used to probe its properties as well as the interaction term in Section~\ref{sec:Open systems}}. Specifically, we consider networks of identical interacting quantum harmonic oscillators where the network structure is generated with Network Geometry with Flavor (NGF). The relation between the eigenvalues of the unweighted Laplacian matrix and the frequencies of the normal modes of the oscillator networks is given and used to connect the spectral properties of the Laplacian matrix with those of the physical network. In Section~\ref{sec:probing} we show how the spectral dimension of the eigenvalues can be estimated from the normal mode frequencies. In Section~\ref{sec:missing}, we consider the impact of small errors and missing values on the estimate and find it to be robust to both. While the normal mode frequencies are possible to probe with an auxiliary system, the normal mode structure can make it difficult to reveal and resolve enough of them; we propose to couple to multiple network nodes with random couplings to avoid this problem. We conclude in Section~\ref{sec:conclusions}.

\section{\label{sec:QNGF}Quantum network geometries}

\subsection{\label{sec:NGF}Network Geometry with Flavor}

NGF \cite{bianconi2016network} is a model of random simplicial complexes, also interpretable as  hyper-graphs.  Simplicial complexes are ideal to model discrete network geometry because they are formed by geometrical building blocks such as triangles, tetrahedra and so on.
Specifically these building block are called simplices. A   $d$-dimensional simplex  include  $d+1$ vertices  forming a  fully connected graph with $d+1$ nodes. A face of a simplex is a lower dimension simplex formed by any proper subset of its nodes. A simplicial complex is formed by simplices that are connected along their faces.  The dimension of a simplicial complex is defined to be the highest dimension of its simplices. 
The NGF are pure $d$-dimensional simplicial complexes, i.e. they are formed from a set of $d$-dimensional simplicies connected along  their $(d-1)$-dimensional faces; a NGF of $d=1$ consists of links connected through their nodes, a NGF of $d=2$ of triangles connected through their links, a NGF of $d=3$ of tetrahedra connected through their triangles and so on.

A $d$-dimensional NGF of $N$ nodes evolves from a single $d$-dimensional simplex by choosing at each timestep a $(d-1)$-dimensional face to which an additional simplex is added, increasing the number of nodes by one. The probability to choose a particular face is affected by the number of simplices already incident with it, as controlled by flavor $s$.  Let $n_\alpha$ be the number of $d$-dimensional simplices incident with face $\alpha$ minus one. The probability $  \Pi_{s}(\alpha)$ to choose face $\alpha$ is given by 
\begin{equation}
    \Pi_{s}(\alpha)=\frac{1}{Z_{s}}(1+sn_\alpha)
\end{equation}
where $Z_{s}$ is
\begin{equation}
    Z_{s}=\sum_\alpha (1+sn_\alpha)
\end{equation}
and the sum is over all $(d-1)$-dimensional faces currently in the growing NGF, ensuring normalization.

A rich variety of simplicial complexes with distinct properties can be generated by adjusting the dimension $d$,  and flavor $s$. In particular a NGF with $s=1$ evolves thanks to an explicit  generalized  preferential attachment mechanism while NGFs with flavor $s=-1,0$ do not obey an explicit preferential attachment rule. Additionally, a $d$-dimensional NGF with flavor $s=-1$ is a discrete manifold; all $(d-1)$-dimensional faces are incident to either $1$ or $2$ $d$-dimensional simplices.

\begin{figure*}[t]
\centering
                \includegraphics[trim=0cm 3.75cm 0cm 3.25cm,clip=true,width=0.95\textwidth]{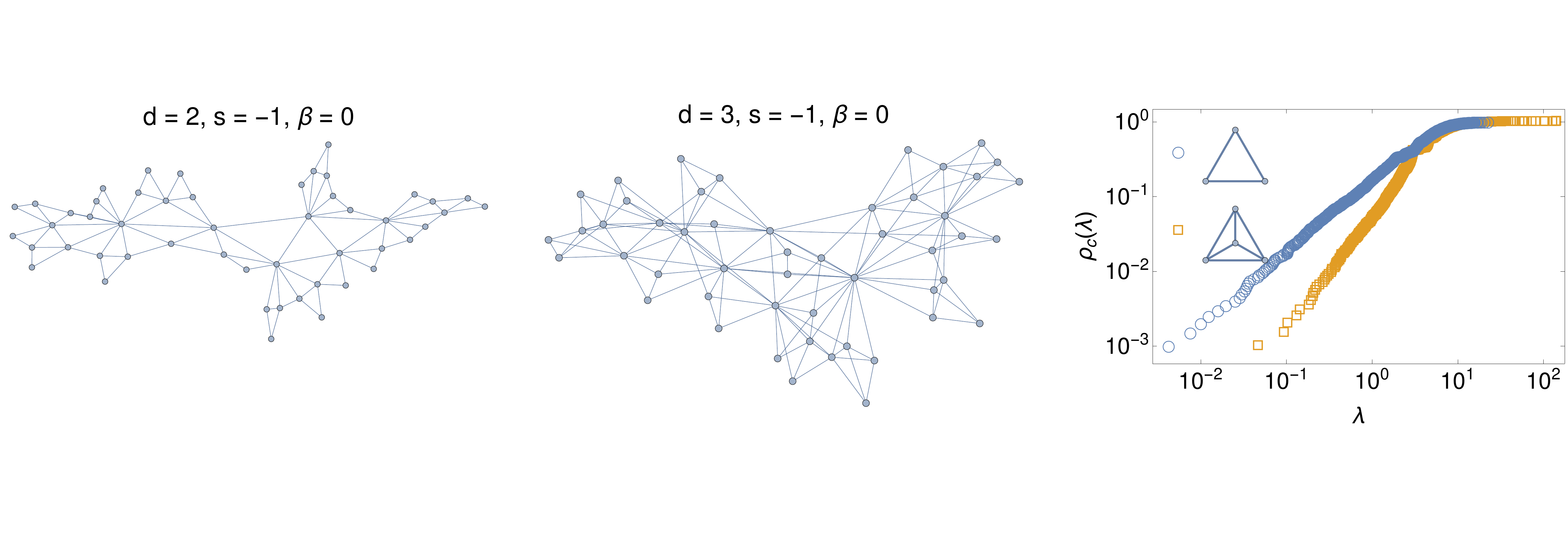}
            \caption{\label{fig:1}
 Examples of NGFs and cumulative distributions of eigenvalues $\rho_c(\lambda)$. The structure is demonstrated with two small ($N=50$) networks. For both each $(d-1)$-dimensional face is incident with at most two $d$-dimensional simplices and each available face is equally likely to attract new simplices during growth. For $d=2$ ($d=3$) the resulting NGF consists of triangles (tetrahedra). The panel on the right shows $\rho_c(\lambda)$ for two NGFs with larger size ($N=2000$) but otherwise same parameters. The power-law scaling for $\lambda\ll 1$ is a sign of a finite spectral dimension $d_S$.
}
      \end{figure*}

The NGFs display an emergent hyperbolic geometry \cite{bianconi2017emergent} as well as a finite spectral dimension $d_S$ \cite{bianconi2020spectral} which is a further indication of their spontaneous geometric character. The spectral dimension is a important spectral property strongly affecting the long time behaviour of the random walk defined on networks. In particular the spectral dimension of a generic (connected) network of $N$ nodes is defined as follows. Let $\mathbf{L}$ be a Laplacian matrix of elements 
\begin{equation}
L_{ij}=\delta_{ij}d_i-a_{ij},
\end{equation}
where $\delta_{ij}$ indicates the Kronecker delta, $d_i$ is the degree of node $i$ and $a_{ij}$ is an element of the adjacency matrix of the network.
Let $\{\lambda_i\}$ indicate  the eigenvalues of the Laplacian matrix $\mathbf{L}$ listed in a non-increasing order, i.e. 
\begin{equation}
0=\lambda_1<\lambda_2\leq \lambda_3\leq \ldots, \leq \lambda_N,
\end{equation} 
where this is valid on any connected network.
If the density of eigenvalues $\rho(\lambda)$ satisfies
\begin{equation}
   \rho(\lambda)=\frac{1}{N}\sum_{i=1}^N\delta(\lambda_i-\lambda)\sim\lambda^{d_S/2-1}
\end{equation}
for  $\lambda\ll1 $, 
with the first non-zero eigenvalue (also called the Fiedler eigenvalue) $\lambda_2$ vanishing in the large network limit as 
\begin{equation}
\lambda_2\sim N^{-2/d_S},
\end{equation}
then $d_S$ is called the spectral dimension of the network. Consequently the cumulative distribution $\rho_c(\lambda)$ scales as
\begin{equation}
\rho_c(\lambda)=\frac{1}{N}\sum_{i=1}^N\theta(\lambda_i-\lambda)\sim\lambda^{d_S/2},    
\label{eq:pclambda}
\end{equation}
for $\lambda\ll 1,$ where $\theta(x)=1$ if $x\geq0$ and $\theta(x)=0$ otherwise. In other words, when the Fiedler eigenvalue of the networks $\lambda_2$ goes to zero in the large network limit, and  $\rho(\lambda)$ and $\rho_c(\lambda)$ has a power-law scaling for small values of $\lambda$, the network has a finite spectral dimension. The spectral dimension for a regular lattice reduces to the Hausdorff dimension of the lattice. In general for networks and fractal geometry, the spectral dimension can be understood as the effective dimension of the networks as probed by a random walker moving on the network.

NGFs  display a finite spectral dimension for every flavor $s\in \{-1,0,1\}$. Interestingly the presence of a finite spectral dimension is robust to modification of the models including non-integers flavors $s=-1/m$ with $m>1$ and the generalization of NGFs to cell-complexes, i.e. discrete topological structures obtained by gluing regular polytopes others than simplices such as hypercubes, orthoplex and so on.

Here for simplicity we focus exclusively on the case $s=-1$. In this case the spectral dimension $d_S$ is an increasing function of $d$.  Examples of both NGF with $d=2$ and $d=3$ flavor $s=-1$  and the scaling of the associated $\rho_c(\lambda)$ are shown in Fig.~\ref{fig:1}. From this figure it is possible to clearly appreciate that the spectral dimension $d_S$ is increasing with increasing values of $d$.

\subsection{\label{sec:QNGFsub}Quantum Network Geometry with Flavor}

In this work we  investigate the properties of a quantum network geometry called  QNGF. The QNGFs are formed  by a set of $N$ quantum harmonic oscillators interacting through the topology described by NGF.
Each quantum harmonic oscillator has  the same  frequency $\omega_0\geq0$, and each pair of connected oscillators is coupled with the same interaction strength $g>0$.
The resulting Hamiltonian reads
\begin{equation}\label{eq:ham}
{H}=\frac{\mathbf{p}^\top\mathbf{p}}{2}+\mathbf{q}^\top\mathbf{A}\mathbf{q},
\end{equation}
where $\mathbf{p}$ and $\mathbf{q}$ are vectors indicating the momentum and position operators of each of the $N$ nodes of the NGF.
The  matrix $\mathbf{A}$ present in the Hamiltonian $H$ can be expressed in term of the Laplacian $\mathbf{L}$ as
\begin{equation}\label{eq:matrixA}
\mathbf{A}=\frac{1}{2}\omega_0^2\mathbf{I}+\frac{1}{2}g\mathbf{L}.
\end{equation}

Since the Hamiltonian~\eqref{eq:ham} is quadratic in position and momentum operators and $\mathbf{A}$ is positive definite, we may find a basis  {of non-interacting} normal modes \cite{tikochinsky1979diagonalization}. To this end let us now define the matrix $\mathbf{U}$ whose columns are the eigenvectors of $\mathbf{A}$, then the diagonal matrix ${\bf D}$ whose diagonal elements are the eigenvalues of ${\bf A}$, is given by
\begin{equation}
{\mathbf D}=\mathbf{U}^\top\mathbf{A}\mathbf{U}.
\label{eq:diagonalization}
\end{equation} 
It follows that by defining new operators for the normal modes as $\mathbf{Q}=\mathbf{U}^\top\mathbf{q}$ and $\mathbf{P}=\mathbf{U}^\top\mathbf{p}$ the Hamiltonian takes a diagonal form where the modes are clearly decoupled, i.e. 
\begin{equation}
H=\frac{\mathbf{P}^\top\mathbf{P}}{2}+\mathbf{Q}^\top\mathbf{D}\mathbf{Q},
\end{equation}
{and their frequencies $\omega_i$ are given by $D_{ii}=\frac{1}{2}\omega_i^2$.}

{We may now relate the eigenvalues of the NGF to the normal mode frequencies of the associated QNGF.}
In fact  from Eq. (\ref{eq:matrixA})  it is clearly evident that the matrix ${\bf A}$ and the Laplacian can be diagonalized on the same basis, i.e. the eigenvectors of ${\bf A}$ are eigenvectors of ${\bf L}$ and vice versa.
Since the eigenvalues of ${\bf A}$ are given by $D_{ii}=\omega_i^2/2$
 Eq. (\ref{eq:matrixA}) implies that the eigenvalues $\lambda_i$ of the Laplacian can be expressed in terms of the frequencies $\omega_i$ as
\begin{equation}\label{eq:lambdaomega}
\lambda_i=\frac{\omega_i^2-\omega_0^2}{g}.
\end{equation}
Therefore for quantum network geometries displaying a finite spectral dimension (like QNGF), we can consider the {cumulative distribution of the normal mode frequencies} $P_c(\omega)$, 
defined as 
\begin{equation}
P_c(\omega) = \frac{1}{N}\sum_i^N\theta(\omega_i-\omega).
\end{equation}
The cumulative distribution of normal model freqquencies $P_c(\omega)$ is related to the cumulative density of the eigenvalues $\rho_c(\lambda)$  of the  Laplacian   by
\begin{equation}
P_c(\omega)= \rho_c((\omega^2-\omega_0^2)/g). 
\end{equation}
In presence of a finite spectral dimension  $d_S$ the cumulative density of eigenvalues $\rho_c(\lambda)$ of the NGF Laplacian obeys  Eq.~\eqref{eq:pclambda}. Therefore it follows that $P_c(\omega)$ scales like 
\begin{equation}
P_c(\omega)\propto(\omega^2-\omega_0^2)^{d_S/2}.
\label{eq:pcomega}
\end{equation}
This relation is key for our work as it implies that the knowledge about the normal mode frequencies reveals information about the spectral dimension $d_S$ of the QNGF.

\subsection{\label{sec:Open systems}   QNGF as a quantum  environment of an open quantum system}

We consider the QNGF as a quantum environment interacting  {weakly} with a single probe formed by an open quantum system. By only considering local observables of the probe our general problem is to study which properties of the QNGF can be inferred. In previous studies of quantum complex networks it has been shown that a similar approach is able to infer several important properties of the quantum networks, e.g.,  {the network spectral density \cite{nokkala2016complex} or degree sequence \cite{nokkala2018local}}. Here we focus specifically on the problem of inferring the spectral dimension of a given QNGF  from the normal mode frequencies $\{\omega_i\}$. In this Section the total Hamiltonian is defined and the principle behind probing the normal mode frequencies is explained. Although here we assume that the network geometry is generated with NGF, our approach is very general and can be applied to any quantum network geometry displaying a finite spectral dimension,  {provided that the Hamiltonian is of the form given by Eq.~\eqref{eq:ham}}. 

The open system is assumed to be an additional quantum harmonic oscillator with a Hamiltonian $H_s$, given by
\begin{equation}
H_s=\frac{p_s^2}{2}+\frac{\omega_s^2q_s^2}{2}
\end{equation}
where  $\omega_s$ is the frequency of the probe and $q_s$ and $p_s$ are the  position and momentum operators respectively. The interaction Hamiltonian is assumed to be of the form 
\begin{equation}
H_I=-q_s\mathbf{k}^\top\mathbf{q},
\end{equation}
where the vector ${\bf k}$ has elements $k_i\geq0$ indicating the interaction strength between the probe and network oscillator $i$. 
 The total Hamiltonian of the  system including both the probe and   the environment is therefore given by 
\begin{equation}
\begin{split}
H_{\textrm{tot}}& = H_s+H+H_I \\
&= \frac{p_s^2}{2}+\frac{\omega_s^2q_s^2}{2}+\frac{\mathbf{p}^\top\mathbf{p}}{2}+\mathbf{q}^\top\mathbf{A}\mathbf{q}-q_s\mathbf{k}^\top\mathbf{q}. 
\label{eq:totalhamiltonian}
\end{split}
\end{equation}
In the bases of the normal modes of the QNGF the total Hamiltonian $H_{\textrm{tot}}$ can be  expressed as  
\begin{equation}
H_{\textrm{tot}}=\frac{p_s^2}{2}+\frac{\omega_s^2q_s^2}{2}+\frac{\mathbf{P}^\top\mathbf{P}}{2}+\mathbf{Q}^\top\mathbf{D}\mathbf{Q}-q_s\mathbf{g}^\top\mathbf{Q},
\label{eq:totalhamiltoniannormalmodebasis}
\end{equation}
where 
\begin{equation}
{\bf g}=\mathbf{U}^\top\mathbf{k}.
\label{eq:g}
\end{equation}
From this expression of the total Hamiltonian it can be appreciated how the eigenvalues of the Laplacian control the physical frequencies affecting the open system while the eigenvectors affect how strong is the open system's coupling to each eigenmode.
Indeed  Eq.~\eqref{eq:g} relates the eigenvectors of the NGF that constitute matrix $\mathbf{U}$ to coupling strengths to the network normal modes. An important special case is the one in which the probe is coupled  to a single node (oscillator). In this case $\mathbf{k}$ has only one non-vanishing element and $\mathbf{g}$  given by Eq. (\ref{eq:g})  is proportional to a single left eigenvector of the Laplacian $\mathbf{L}$.
Since according to   Eq.~(\ref{eq:g}) ${\bf g}$ is linear in ${\bf k}$, and since ${\bf k}$ can be written as 
\begin{equation}
{\bf k}=\sum_{i=1}^Nk_i{\bf e}^i
\end{equation}
where ${\bf e}^i$ is the vector  of elements $e^i_j='\delta_{ij}$, it follows that in the general case the vector ${\bf g}$ can be interpreted as a linear combination of the interaction strength resulting from coupling to single nodes of the network.

Since the Hamiltonian is quadratic, the dynamics can be both solved and simulated efficiently when the initial states of the network and the probe are Gaussian states \cite{bartlett2002efficient}; for an explicit example see, e.g., \cite{nokkala2017non}.

The considered local observable is the expected value of the probe excitations, $\langle a^\dagger a\rangle$, where $a^\dagger$ and $a$ are the probe creation and annihilation operators, respectively. This quantity is proportional to the energy of the probe. {Let $\langle a(0)^\dagger a(0)\rangle$ be the expected value of the probe initial excitations, and suppose it evolves with the network according to} total Hamiltonian $H_{\textrm{tot}}$ for time $t$, taking probe excitations to $\langle a(t)^\dagger a(t)\rangle$. The probe is then decoupled from the network and its excitations are measured, allowing the determination of the quantity $\Delta n=|\langle a(t)^\dagger a(t)\rangle-\langle a(0)^\dagger a(0)\rangle|$. It is well-known that an open system weakly interacting with a bosonic heat bath must be in resonance with the bath to exchange information and excitations with it. In the present case this means that for sufficiently weak interaction, $\Delta n\gg0$ implies that $\omega_S\approx\omega_i$ for some $i\in\{1,2,\ldots,N\}$. The principle is then to repeat the protocol for many different values of $\omega_S$; a large $\Delta n$ indicates that a normal mode frequency is in the vicinity of the used $\omega_S$.

In the following Section we focus on the estimation of $d_S$ assuming that we have acquired the full set of normal mode frequencies $\{\omega_i\}$ from probing them. We will discuss only later, in section ~\ref{sec:missing}, the more realistic situation in which we are able to obtain only some approximate values of normal mode frequencies $\{\omega_i\}$ and briefly consider general strategies to infer $\{\omega_i\}$ from $\Delta n$.

\section{\label{sec:probing}Probing the spectral dimension}

 {Given the normal mode frequencies $\{\omega_i\}$ of a QNGF, the objective is to estimate its spectral dimension $d_S$.} Our strategy is to find the least-squares fit of $\{\omega_i\}$ into a model with $d_S$ as a fitted variable. We begin by recasting the power-law scaling in relation~\eqref{eq:pcomega} to a linear one by considering the logarithm of both sides, namely
\begin{equation}\label{eq:pcomegalog}
    \log(P_c(\omega))\sim(d_S/2)\log(\omega^2-\omega_0^2).
\end{equation}
The fitting is now amenable to ordinary least squares and the best fit can be found applying the Moore-Penrose pseudoinverse \cite{penrose1956best}. While fitting to a linear model requires to know the bare frequency $\omega_0$ it is not necessary to know it \textit{a priori} since it will coincide with the smallest normal mode frequency; this follows from the well-known property $\textrm{min}(\{\lambda_i\})=0$ of Laplace eigenvalues and from Eq.~\eqref{eq:lambdaomega}. 

At this point we face two additional problems. First, the normal mode frequencies are expected to behave according to relations~\eqref{eq:pcomega} and \eqref{eq:pcomegalog} only up to some upper limit in frequencies; therefore to estimate $d_S$ we should have a way to truncate the frequencies appropriately. Second, we would like to be able to say something about the quality of the estimated value and in particular be able to compare different estimates resulting from different points of truncation. To address the first point we consider the goodness-of-fit between the linear model and actual data derived from $\{\omega_i\}$ for different points of truncation and choose the optimal value; specifically, we consider the coefficient of determination $R^2$ \cite{devore2011probability}. For the second, we consider the $95 \%$ confidence intervals. Further details are given in the \ref{app:fitting}.

Remarkably, the estimate is independent of the values of $\omega_0$ and $g$. On the one hand this is because the value of $\omega_0$ can in principle be probed exactly which facilitates the use of units where it becomes irrelevant. On the other hand $g$ can only scale the L.H.S. of Eqs.~\eqref{eq:pcomega} and \eqref{eq:pcomegalog}; this does not affect the goodness-of-fit and therefore not the values of $R^2$ nor the confidence intervals as a function of the index $i$ which determines the cutoff frequency $\omega_i$.

\begin{figure*}[t]
\centering
                \includegraphics[trim=0.15cm 0.35cm 0.25cm 0.25cm,clip=true,width=0.65\textwidth]{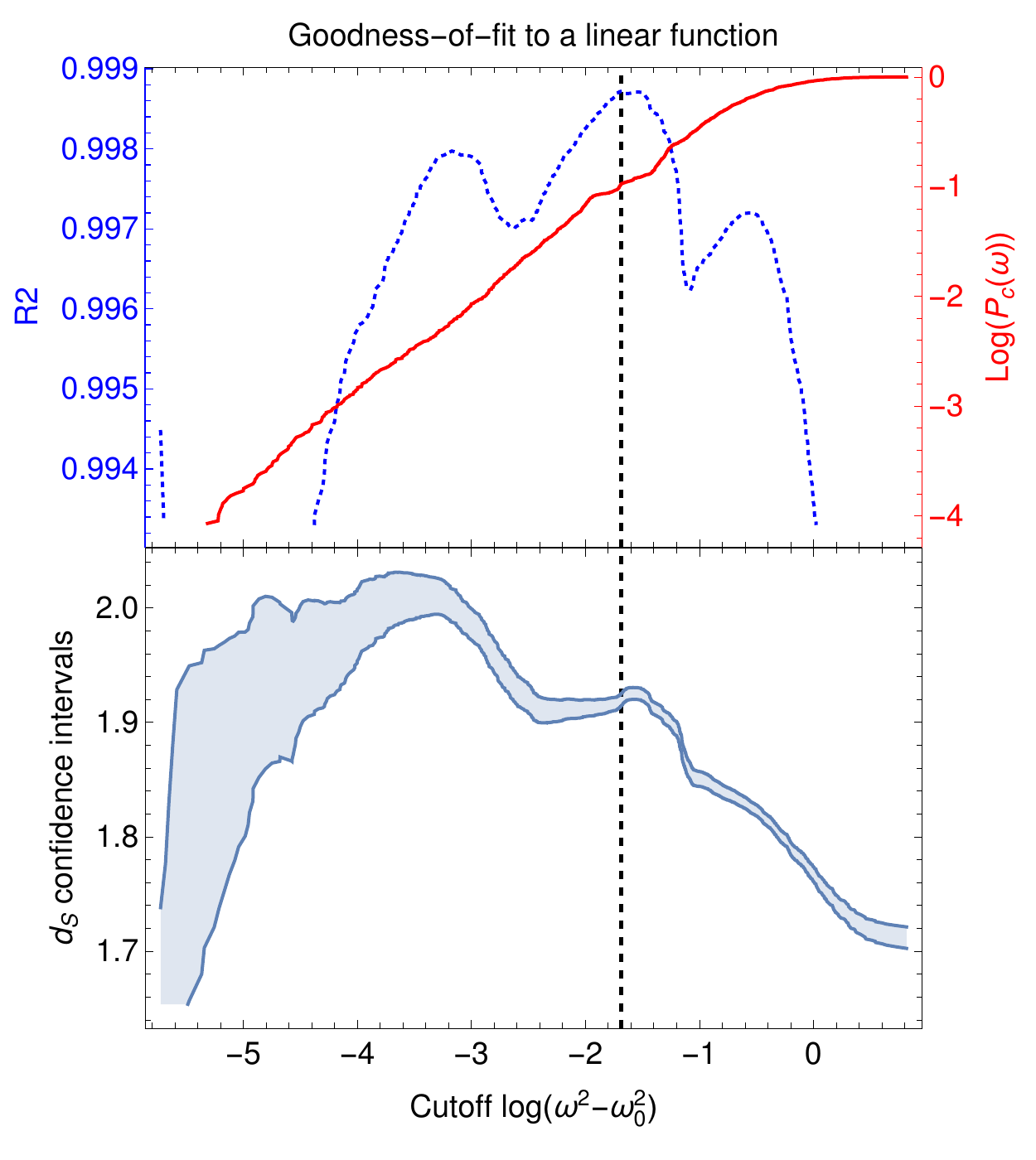}
            \caption{\label{fig:2}
Estimation of the spectral dimension by a linear fit. Top panel: the transformed $P_c(\omega)$ (solid line) is shown together with the coefficient of determination $R^2$ (dotted line). The latter gives the goodness-of-fit of a linear model fitted to data truncated at the position on the horizontal axis. Bottom panel: the $95 \%$ confidence intervals for the estimated value of $d_S$, given by the slope of the linear model fitted to truncated data. The optimal value of $R^2$ is shown by the vertical dashed line in both panels. The QNGF has 2000 nodes with $\omega_0=0.25$ interacting with a coupling strength $g=0.1$, while $d=2$,  and $s=-1$.
}
      \end{figure*}

A specific example of this procedure is shown  in Fig.~\ref{fig:2} where we consider a QNGF whose  underlying topology is NGF of $N=2000$  nodes, with $d=2$ and $s=-1$. As explained before, the particular values of $\omega_0=0.25$ and $g=0.1$ do not affect the estimate. In the top panel the logarithm of $P_c(\omega)$ is shown as a function of $\log(\omega^2-\omega_0^2)$ together with $R^2$ as a function of the point of truncation; higher values indicate a better agreement between the fit and the data. Here $d_S$ is proportional to the slope of the linear model. Below the corresponding confidence intervals for $d_S$ are shown. The optimal value of $R^2$, showing the optimal point to truncate the frequencies, is shown by the vertical dashed line. There is clear correlation between the behaviour of $R^2$ and the confidence intervals, with higher values leading to a smaller interval which we interpret as a more reliable estimate for $d_S$.

\begin{figure}[t]
\centering
                \includegraphics[trim=0.15cm 0.15cm 0cm 0cm,clip=true,width=0.65\textwidth]{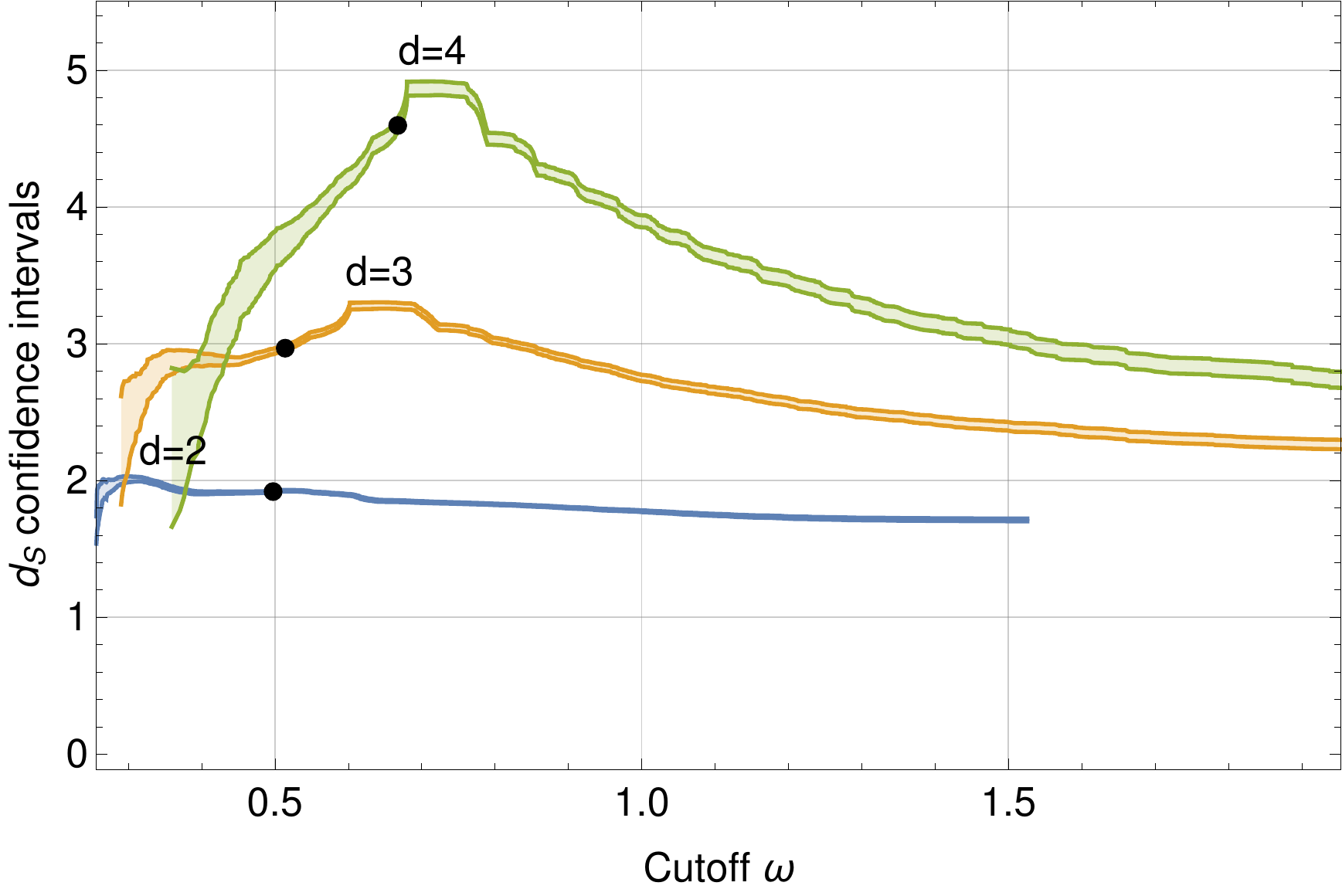}
            \caption{\label{fig:3}
Comparison of $95 \%$ confidence intervals for three QNGF with $d=2,3,4$ and $s=-1$ as a function of the cutoff frequency where data is truncated before fitting. The black dots show the estimated value of $d_S$ at the optimal value of the goodness-of-fit measure $R^2$. All parameter values aside from $d$ are as in Fig.~\ref{fig:2}.
}
      \end{figure}

We compare the confidence intervals of Fig.~\ref{fig:2} with those of two other networks which have otherwise the same parameters but have dimensions $d=3$ and $d=4$, shown as a function of the truncation frequency in Fig.~\ref{fig:3}. Here it can be better appreciated that the estimates are not particularly sensitive to small changes in the truncation frequency. On the other hand, the results suggest that this robustness decreases for increasing values of the spectral dimension $d_S$.

Before concluding the Section, we point out the possibility to fit Eq.~\eqref{eq:pcomega} directly to a nonlinear model. Similar arguments can be employed to show that such an estimate is also independent of the values of $\omega_0$ and $g$, however in general finding a nonlinear least-squares fit is more challenging and in particular the available algorithms, such as Levenberg-Marquardt method \cite{levenberg1944method}, nonlinear conjugate gradient method \cite{polak1969note,polyak1969conjugate} and quasi-Newton methods \cite{dennis1977quasi}, do not necessarily converge to a global optimum. This can lead to, e.g., jumps from one local optimum to another as the cutoff frequency is varied. We also remark that there are many other goodness-of-fit measures that could be employed; in the present case numerical experiments suggest that they all lead to similar results.

\section{\label{sec:missing}Robustness to missing or noisy normal mode frequencies}

As outlined  {in Section~\ref{sec:Open systems}}, normal mode frequencies $\{\omega_i\}$ can be  {inferred} from the dynamics of an open quantum system, or probe, weakly interacting with the QNGF, by performing a frequency sweep in the low frequency regime and tracking {, e.g., the change in probe excitations}. In practice, only a finite set of values for $\omega_S$ can be considered and consequently the probed values of $\{\omega_i\}$ will not be exact. On the other hand, the cumulative distribution $P_c(\omega)$ can be expected to be very robust to small errors. Provided that we are able to resolve different normal mode frequencies these errors would be smaller than the difference between adjacent normal mode frequencies, leaving $P_c(\omega)$ almost unaffected. We have checked that even a very pessimistic case of i.i.d. relative error up to $5 \%$ leaves all results shown in Figs.~\ref{fig:2} and \ref{fig:3} virtually the same. 

Rather than finding accurate values for $\{\omega_i\}$, the challenge is to find values for them at all. In the extreme case where $g_i=0$ for some index $i$ in Eq.~\eqref{eq:g} it is impossible to learn the corresponding normal mode frequency $\omega_i$ since the probe is decoupled from this normal mode. Since $\mathbf{g}$ is a linear combination of the left eigenvectors of matrix $\mathbf{L}$ of the NGF, this is highly unlikely to occur as long as $\mathbf{k}\neq\mathbf{0}$. What is to be expected, however, is that normal modes with frequencies in close proximity to each other interact with the probe with very different strengths. In such a case the impact on probe dynamics is dominated by the normal modes with a stronger coupling strength, making it almost impossible to reveal the other nearby frequencies. It is also difficult to resolve frequencies that are close and interact with the probe with similar strengths, in which case many frequencies might be mistakenly counted as a single frequency.

\begin{figure}[t]
\centering
                \includegraphics[trim=0cm 0cm 0cm 0cm,clip=true,width=0.95\textwidth]{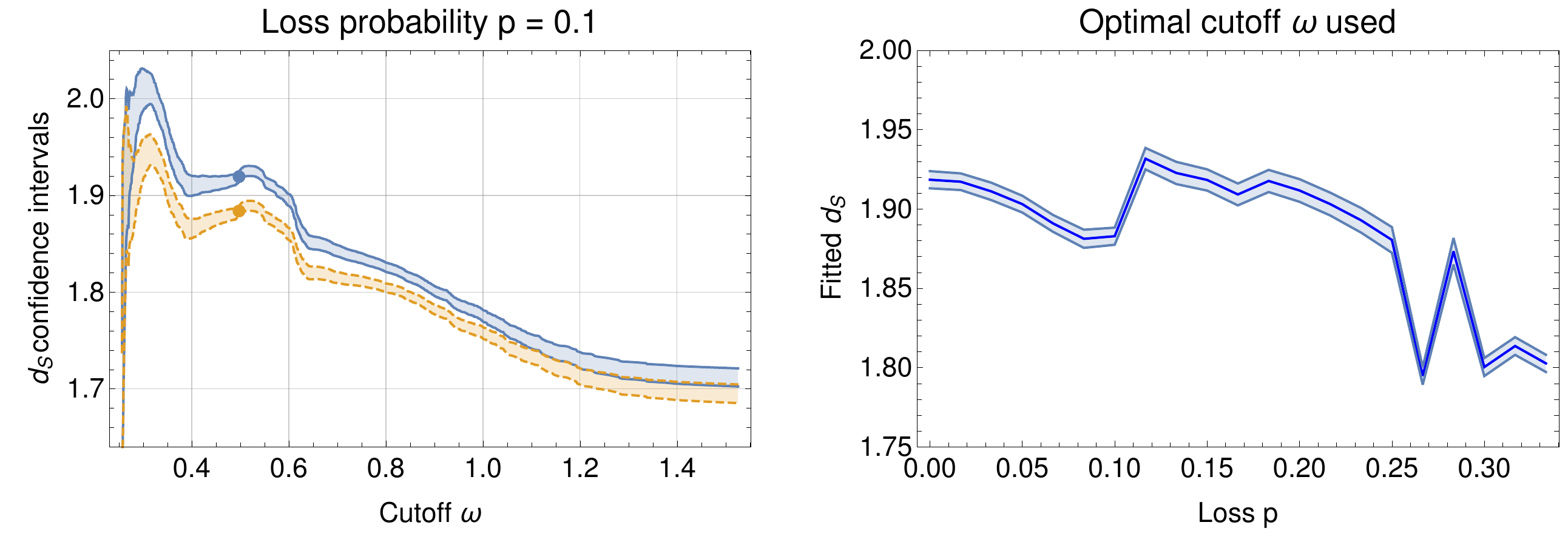}
            \caption{\label{fig:4}
Left: comparison of confidence intervals derived from complete (solid lines) and partial (dashed lines) data. In the latter case a random sample of $10 \%$ of the eigenfrequencies is discarded before estimation. The dots show the fitted value of $d_S$ at the optimal value of the goodness-of-fit measure $R^2$. Right: the estimated value of $d_S$ (middle line) and confidence intervals (outer lines) when the optimal cutoff frequency is used as a function of the probability $p$ of missing data. The network parameters are as in Fig.~\ref{fig:2}.
}
      \end{figure}

We consider {the case where each normal mode frequency is equally likely to be missing} in Fig.~\ref{fig:4}. In the left panel we fix the probability to miss a normal mode frequency to $p=0.1$ and compare the confidence intervals when full and partial set of normal mode frequencies is available, indicated by solid and dashed lines, respectively. The fitted value of $d_S$ at the optimal cutoff frequency is shown by the dot. In the right panel we show the estimated value of $d_S$ which maximizes the goodness-of-fit against probability $p$. It can be seen that neither the confidence intervals nor the optimal estimate for $d_S$ are particularly sensitive to uniformly missing frequencies. As $p$ increases the trend seems to be that the optimal estimate decreases, however the result is still reasonably close to original even {at $p=0.3$}.

In a probing scenario missing frequencies might not be uniformly distributed, while in some cases a frequency might be mistakenly interpreted as a normal mode frequency. To further investigate these challenges we  simulated a simple proof-of-concept scheme where the QNGF spectrum is swept using equidistant values of probe frequency and the change in probe excitations $\Delta\langle n\rangle=|\langle n(0)\rangle-\langle n(t)\rangle|$ from initial to final time $t$ is considered as a function of probe frequency $\omega_S$. Prominent local maxima indicate resonance, i.e. that a normal mode frequency is close. A general purpose non-adaptive algorithm is employed to find these maxima and $d_S$ is estimated from the associated values of probe frequency. The considered network is a small QNGF with $N=50$, $d=2$, and $s=-1$. We took the initial states to be vacuum for the network and squeezed vacuum with squeezing parameter $r=2.5$ for the probe. Further details about the dynamics, the frequency sweeps and the algorithm used to find the peaks are given in \ref{app:dynamics}, \ref{app:sweeps} and \ref{app:quantum}, respectively.

\begin{figure}[t]
\centering
                \includegraphics[trim=0cm 0cm 0cm 0cm,clip=true,width=0.95\textwidth]{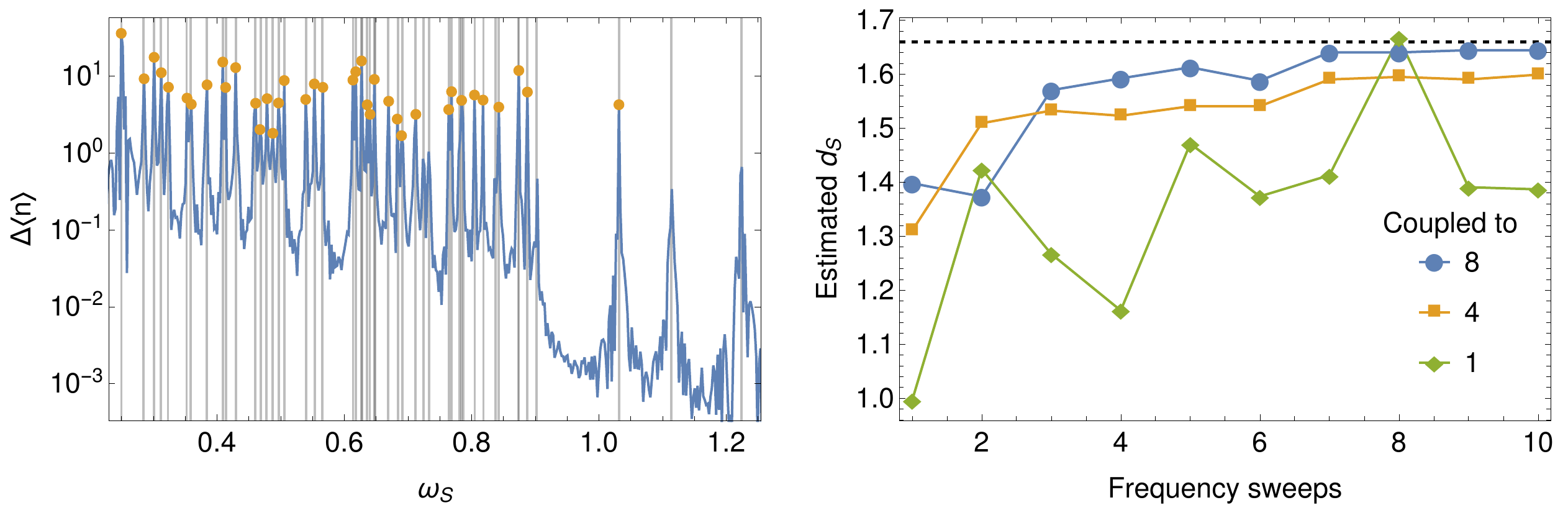}
            \caption{\label{fig:5}
Example of spectral dimension probing with a small QNGF ($N=50$). Left: change in probe excitations $\Delta\langle n\rangle$ as a function of probe frequency $\omega_S$ averaged over 10 sweeps when the probe is coupled to 8 randomly selected network nodes. The normal mode frequencies are indicated by vertical lines and probed values by dots. Right: estimated $d_S$ when coupling the probe to an indicated number of randomly selected network nodes. In both cases a new random selection is made for each sweep. The initial state is squeezed vacuum for the probe and vacuum for the network. See main text and \ref{app:quantum} for details.
}
      \end{figure}

The results are shown in Fig.~\ref{fig:5}. In the {right} panel an example of the behaviour of $\Delta\langle n\rangle$ is shown. Here the shown values are averaged over 10 frequency sweeps. For each sweep the probe is coupled to 8 randomly chosen network nodes. The vertical lines indicate the values of $\{\omega_i\}$ and the dots are their probed values, found by identifying the peaks. As can be seen, the peaks align quite well with $\{\omega_i\}$. Even though the majority of normal mode frequencies are found, some are missed, including rather obvious cases in the high frequency regime. While indeed there is a local maximum in the vicinity of every normal mode frequency it will be seen that avoiding false positives is important for probing $d_S$. Consequently very cautious settings for the peak finding algorithm are used in the entire spectrum; further improvements could be expected by using a custom algorithm optimized for the task at hand.

In the left panel we compare $d_S$ estimated from original $\{\omega_i\}$ (dashed vertical line) to values estimated from probed normal mode frequencies as a function of the number of performed frequency sweeps and when a different amount of randomly selected network nodes is directly interacting with the probe. The results are not sensitive to cutoff frequency in the vicinity of its optimal value, which is used throughout. It can be seen that the accuracy tends to improve with the number of sweeps. This is to be expected since each set of network nodes corresponds to a different linear combination of the associated left eigenvectors of $\mathbf{L}$. Different sweeps therefore change which normal modes are interacting too weakly to be distinguished. While the trend is not as strong, it also seems to be the case that it is better to couple to multiple network nodes than just one. This is actually connected to the number of false positives which tends to be highest when coupling to a small number of network nodes; in terms of correctly identified $\{\omega_i\}$ alone there is no evident difference between the three cases, which all saturate to approximately $80\%$ of $\{\omega_i\}$ with the used algorithm and its settings.

It may be asked what role the initial conditions play in the final result. As a matter of fact, a simple analytical argument suggests that when the network is in vacuum the estimate can easily be made independent of the probe initial excitations by scaling $\Delta\langle n\rangle$. Furthermore, the probe initial state will not change the estimate either. The details are given in \ref{app:states} where we make the argument and verify in Fig.~\ref{fig:6} that it holds to a good approximation by simulating the spectral dimension probing for many different probe initial states and initial excitations. We also confirm that probing remains robust when the network is in a thermal state of some finite temperature, which suggests robustness for other network states as well, as argued in the Appendix. This all assumes that sufficiently many measurements are done to deduce the value of $\Delta\langle n\rangle$. While it is conceivable that using non-classical states for the probe can optimize the number of measurements, such considerations are beyond the scope of the present work.

In general, the challenges revolve around resolving a sufficiently large fraction of $\{\omega_i\}$ while avoiding false positives, which in practice requires large differences of $\Delta\langle n\rangle$ when in and out of resonance. The difficulty increases with the density of $\{\omega_i\}$ and therefore with $N$ and $d$, while it tends to decrease with longer interaction times and weaker interaction. Multiple frequency sweeps and coupling to multiple network nodes can be expected to remain beneficial. False positives in particular seem to cause a loss in the rise and fall behaviour of the goodness-of-fit measure, moving the maximum value to high frequency regime. This could potentially be used to eliminate them.

\section{\label{sec:conclusions}Conclusions}

Finite spectral dimension is characteristic of network geometries. A very general mathematical framework for generating a large variety of  network geometries is the ``Network Geometry with Flavor" model. The simplicial complexes generated by this model obey   simple stochastic rules  yet from these simple rules network geometries emerge spontaneously. Here we have proposed the model QNGF which is formed by a set of quantum harmonic oscillators interacting  through a hyperbolic simplicial network geometry given by NGF.  {Using} the relation between the normal mode frequencies of QNGFs and the eigenvalues of the Laplacian of NGFs we show that the spectral dimension of NGF can be probed with an open quantum system. The obtained estimate of the spectral dimension is independent of the bare frequency of the network oscillators and the strength of the couplings between them and is not sensitive for the choice of the cutoff frequency, especially when the spectral dimension is small. Our estimate of the spectral dimension  is formed from the normal mode frequencies of the network, which are in principle possible to deduce from the reduced dynamics of the probe. In practice some deviation from the exact values and incompleteness of the set of frequencies can be expected. Our results show that the estimate remains robust to small deviations and uniformly missing frequencies. As a further proof-of-concept we simulated probing of the normal mode frequencies and the spectral dimension for a small network. The results reveal that misinterpreting a frequency as a normal mode frequency can be just as harmful for the estimation as not finding one, and in particular the two effects do not tend to cancel each other out. To reduce both errors simultaneously we propose to perform multiple frequency sweeps where the probe is coupled to multiple randomly selected network oscillators. Indeed, in this way the estimate is close to that of the ideal case. We expect the proposed method to be useful also for probing just the normal mode frequencies.

The QNGF provides a very flexible benchmark model to investigate quantum network geometry.  In particular, this allows us to explore the robustness of the results in higher dimension, by comparing the results obtained on NGF evolving according to the same rules but having different dimension $d$ of their building blocks and different spectral dimension $d_S$. We show  that  probing the spectral dimension can be performed on  QNGF formed by simplices of different dimension $d$ although we achieved higher  accuracy of the results for lower dimensions. Interestingly the approach proposed in this work and tested  on QNGF can be applied  to arbitrary quantum network geometries  displaying  a finite spectral dimension.

\section{Data availability}

The data that support the findings of this study are available upon reasonable request from the authors.

\ack

J.N. acknowledges funding from the Spanish State Research Agency, through the Severo Ochoa and Mar\'ia de Maeztu Program for Centers and Units of Excellence in R\&D (MDM-2017-0711) and thanks Turku Centre for Quantum Physics for the hospitality.

\appendix

\section{\label{app:fitting}Used goodness-of-fit measure and determination of parameter confidence intervals}

The coefficient of determination $R^2$ is a measure of how well the predictions of the fit approximate the data points, defined as
\begin{equation}
    R^2=\frac{\sum_i(f_i-\bar{y})^2}{\sum_i(y_i-\bar{y})^2}
\end{equation}
where $f_i$ are predicted values and $\bar{y}$ is the mean value of the data points. Notice that the denominator is proportional to the variance of the data. Normally $0\leq R^2\leq 1$, where $R^2=0$ indicates that the variance in the data cannot be explained at all by the fit while $R^2=1$ indicates that the variance can be explained perfectly.

Besides $R^2$, we also considered adjusted $R^2$ \cite{theil1961economic}, Bayesian information criterion \cite{schwarz1978estimating} and Akaike information criterion with \cite{cavanaugh1997unifying} and without \cite{akaike1974new} correction for small sample sizes. Unsurprisingly, adjusted $R^2$ behaves very similarly to $R^2$, while the latter three have very similar behaviour with each other but differ from that of $R^2$ and adjusted $R^2$. Specifically, for them the optimal value seems to appear at somewhat higher frequencies than for $R^2$ and adjusted $R^2$, which in turn tends to correspond to a bit smaller estimate for $d_S$.

The considered parameter confidence intervals are defined to be intervals of values such that they include the true value $95 \%$ of the time. To determine the intervals it is assumed that the error in the parameters is normally distributed. Let $\alpha=1-0.95$, $n$ the number of data points and $p=2$ the number of parameters. Then the intervals are
\begin{equation}
    d_S\pm t(1-\alpha/2,n-p)\textrm{SE}(d_S)
\end{equation}
where $d_S$ is the estimated spectral dimension, $t(1-\alpha/2,n-p)$ is the $100(1-\alpha/2)$ percentile of Student's $t$ distribution with $n-p$ degrees of freedom and $\textrm{SE}(d_S)$ is the standard error of the estimated spectral dimension. $\textrm{SE}(d_S)$ is determined from the parameter covariance matrix $\hat{\sigma}^2(\mathbf{X}^\top\mathbf{W}\mathbf{X})^{-1}$ where $\hat{\sigma}^2$ is the variance estimate, $\mathbf{X}$ the design matrix and $\mathbf{W}$ the diagonal matrix of weights. The diagonal elements of the parameter covariance matrix are squares of parameter standard errors.

\section{\label{app:dynamics}Determination of the open system dynamics}

Simulation of the dynamics is considered in Sec.~\ref{sec:missing}. Throughout, we work in such units that the reduced Planck constant $\hbar=1$ and the Boltzmann constant $k_B=1$. Oscillators have unit mass.

The Gaussian states considered in this work are a paradigmatic class of states in continuous-variable quantum information science, and may be defined as the states determined completely by the second and first moments of the momentum and position operators \cite{ferraro05gaussian,adesso2014continuous}. Such states are conveniently described by their covariance matrix $\bm{\sigma}$. Consider the Hamiltonian $H_{\textrm{tot}}$ in the basis of network normal modes, given by Eq.~\eqref{eq:totalhamiltoniannormalmodebasis}. Let $\mathbf{X}=\{Q_1,Q_2,\ldots,Q_N,q_s,P_1,P_2,\ldots,P_N,p_s\}$. In this basis the covariance matrix is
\begin{equation}
    \bm{\sigma}_{ij}=\frac{1}{2}\langle[\mathbf{X}_i,\mathbf{X}_j]_+\rangle-\frac{1}{2}[\langle\mathbf{X}_i\rangle,\langle\mathbf{X}_j\rangle]_+,
    \label{eq:covariancematrix}
\end{equation}
where the angle brackets denote an expectation value over the state and $[\mathbf{X}_i,\mathbf{X}_j]_+=\mathbf{X}_i\mathbf{X}_j+\mathbf{X}_j\mathbf{X}_i$ is the anticommutator.

As a Hamiltonian quadratic in momentum and position operators, $H_{\textrm{tot}}$ preserves the Gaussian character of the state. Consequently the evolution it induces is completely captured by the evolution of the covariance matrix. Let $\bm{\sigma}(0)$ be the initial form of the covariance matrix of Eq.~\eqref{eq:covariancematrix} and let the probe and network interact for time $t$, taking the covariance matrix to $\bm{\sigma}(t)$. Then the initial and final forms of the total covariance matrix are related as
\begin{equation}
    \bm{\sigma}(t)=\mathbf{S}(t)\bm{\sigma}(0)(\mathbf{S}(t))^\top,
    \label{eq:dynamics}
\end{equation}
where the matrix $\mathbf{S}(t)$ is induced by the Hamiltonian $H_{\textrm{tot}}$ and the interaction time $t$.

The explicit form of $\mathbf{S}(t)$ is easily found by using the analytic solution for the equations of motion of non-interacting oscillators. To apply it, the total system of network and probe is diagonalized, propagated in the diagonal basis, and taken back to the original basis. This corresponds to decomposing $\mathbf{S}(t)$ as
\begin{equation}
    \mathbf{S}(t)=\begin{pmatrix}
\mathbf{O} & 0\\
0 & \mathbf{O}
\end{pmatrix} \mathbf{S}_{\textrm{diag}}(t) \begin{pmatrix}
\mathbf{O} & 0\\
0 & \mathbf{O}
\end{pmatrix}^\top,
\end{equation}
where $\mathbf{O}$ is the orthogonal matrix diagonalizing $H_{\textrm{tot}}$, akin to $\mathbf{U}$ in Eq.~\eqref{eq:diagonalization}, and $\mathbf{S}_{\textrm{diag}}(t)$ propagates the covariance matrix of non-interacting oscillators. Let us define the following diagonal matrices: $(\bm{\Delta}_{\omega})_{ii}=\omega_i$, $(\bm{\Delta}_{\textrm{cos}})_{ii}=\cos(\omega_i t)$ and $(\bm{\Delta}_{\textrm{sin}})_{ii}=\sin(\omega_i t)$, where $\omega_{N+1}=\omega_s$. Now
\begin{equation}
    \mathbf{S}_{\textrm{diag}}(t)=\begin{pmatrix}
\bm{\Delta}_{\textrm{cos}} & \bm{\Delta}_{\omega}^{-1}\bm{\Delta}_{\textrm{sin}}\\
-\bm{\Delta}_{\omega}\bm{\Delta}_{\textrm{sin}} & \bm{\Delta}_{\textrm{cos}}
\end{pmatrix}.
\end{equation}

As an example initial state for the network we may consider the stationary state of the network Hamiltonian $H$, i.e. a thermal state of some temperature $T$---in the simulations we take $T=0$ unless stated otherwise. The corresponding covariance matrix is diagonal in the basis of network normal modes and has elements
\begin{equation}
\langle Q_i^2\rangle=\frac{1}{2\omega_i}\left(2n_i+1\right), \quad \langle P_i^2\rangle=\frac{\omega_i}{2}\left(2n_i+1\right),
\label{eq:networkinitialstate}
\end{equation}
where $n_i=(\exp(\omega_i/T)-1)^{-1}$.

The initial state of the probe is taken to be a squeezed state, i.e. a state where the second moment in one quadrature is lowered below that of vacuum, at the expense of increasing the second moment in the other quadrature. In the simulations we consider squeezing of the momentum. The corresponding covariance matrix is also diagonal, with elements
\begin{equation}
\langle q_s^2\rangle = \frac{1}{2\omega_s}\exp(2r), \quad \langle p_s^2\rangle = \frac{\omega_s}{2}\exp(-2r),
\label{eq:probeinitialstate}
\end{equation}
where $r$ is the magnitude of squeezing. With this, all of the non-vanishing elements of the initial form of the covariance matrix $\bm{\sigma}$ in Eq.~\eqref{eq:covariancematrix} are fixed. The probe observable we consider is
\begin{equation}
\begin{split}
\langle a^\dagger a\rangle& = \frac{\omega_s}{2}\langle q_s^2\rangle+\frac{1}{2\omega_s}\langle p_s^2\rangle-\frac{1}{2} \\
&= \frac{\omega_s}{2}\bm{\sigma}_{N+1,N+1}+\frac{1}{2\omega_s}\bm{\sigma}_{2N+2,2N+2}-\frac{1}{2},
\label{eq:probeexcitations}
\end{split}
\end{equation}
namely the probe excitations. It is proportional to the probe energy $\langle H_s\rangle$. With the initial sate of the probe defined by Eq.~\eqref{eq:probeinitialstate}, we get the initial value $\langle a^\dagger a\rangle=\sinh^2(r)$, while the value at time $t$ is recovered by substituting $\bm{\sigma}(t)$ of Eq.~\eqref{eq:dynamics} into Eq.~\eqref{eq:probeexcitations}.

\section{\label{app:sweeps}Frequency sweeps}

In Sec.~\ref{sec:missing} the quantity $\Delta n=|\langle a(t)^\dagger a(t)\rangle-\langle a(0)^\dagger a(0)\rangle|$ is considered when frequency sweeps are made. The evaluation of $\langle a^\dagger a\rangle$ is covered in \ref{app:dynamics}, while here the protocol for frequency sweeps is explained.

A single value for $\Delta n$ is acquired as follows.
\begin{enumerate}
    \item the value for probe frequency $\omega_S$ is chosen
    \item the probe is prepared to the initial state with excitations $\langle a(0)^\dagger a(0)\rangle$
    \item the interaction Hamiltonian $H_I$ is switched on for a time $t$
    \item the probe excitations are measured
    \item the process is repeated as necessary to get an accurate value for $\langle a(t)^\dagger a(t)\rangle$ 
\end{enumerate}
In the simulations the state of the network resets with the state of the probe, but in general the results can be expected to be similar as long as there is an energy difference between the probe and the network. To perform a frequency sweep, the above protocol is repeated for many different values of $\omega_S$, resulting in a list of values for $\Delta n$.

In Sec.~\ref{sec:missing} both single and multiple frequency sweeps are considered. In the latter case a frequency sweep is repeated for many different realizations of the interaction Hamiltonian $H_I$, resulting in an array of values for $\Delta n$. This array is reduced to a list of values by averaging over different $H_I$. When choosing network nodes to couple the probe to, each node is selected with equal probability. The selections are independent, consequently overlap between sets for different sweeps may happen.

\section{\label{app:quantum}Normal mode frequency probing}

Here it is explained how the results shown in Fig.~\ref{fig:5} were made. Details about the determination of the open system dynamics and the value of the quantity $\langle a^\dagger(t)a(t)\rangle$ are given in \ref{app:dynamics} while the protocol for frequency sweeps is covered in \ref{app:sweeps}.

The considered network consists of identical quantum harmonic oscillators with a bare frequency of $\omega_0=0.25$ interacting with uniform coupling strengths $g=0.1$. The probe couples to the network in such a way that $\sum_i k_i=0.05g$ while all non-vanishing elements $k_i$ have the same magnitude. At $t=0$ the state of the probe is squeezed vacuum, defined by Eq.~\eqref{eq:probeinitialstate}, with squeezing parameter $r=2.5$, while the network is in vacuum, defined by Eq.~\eqref{eq:networkinitialstate} with $T=0$.

For all shown results, the interaction time in $\Delta\langle n\rangle=|\langle a^\dagger(t)a(t)\rangle-\langle a^\dagger(0)a(0)\rangle|$ is $t=40000\omega_s$, while 500 equidistant values in the closed interval $[0.9\omega_0,1.1\omega_N]$ are used for $\omega_s$. Here $\omega_N$ is the largest normal mode frequency. As explained in \ref{app:sweeps}, the frequency sweeps result in 500 values for $\Delta\langle n\rangle$.

To find the normal mode frequencies from the values of $\Delta\langle n\rangle$, we employ the following algorithm. We perform a Gaussian blurring on the data up to scale  {$\sigma=0.55$}. Out of all of the local maxima of the blurred data, we select any that have a minimum sharpness of $1$ and a value at least $0.1$. In other words, the data is first convolved with a Gaussian kernel of standard deviation $\sigma$. The surviving maxima are chosen if they have a negative second derivative greater in magnitude than $1$ and a value at least $0.1$. The second derivative is estimated by padding the data by a single repetition of both the first and last element and convolving it with the kernel $\{1,-2,1\}$. The role of blurring is to smooth out weak peaks, and sharpness and minimum value further restrict which surviving maxima are selected.

The spectral dimension is estimated from the found normal mode frequencies exactly the same way as in Sec.~\ref{sec:probing}.

\section{\label{app:states}Impact of probe and network initial states}

If the network is initially in its ground state and the probe is not, then changing the initial state of the probe should have negligible impact on the estimated spectral dimension. This can be understood from the following analytic expression of $\langle a^\dagger(t)a(t)\rangle$ (Eq.~$(15)$ in \cite{maniscalco2004simulating}), derived in the infinite size, weak coupling and Markovian limits and assuming a thermal state of some temperature $T$ for the network. It reads
\begin{equation}
    \langle a^\dagger(t)a(t)\rangle=e^{-\Gamma t}\langle a^\dagger(0)a(0)\rangle+(1-e^{-\Gamma t})n_{th}(\omega_S),
\label{eq:excitations}
\end{equation}
where $n_{th}(\omega)=(\exp(\omega/T)-1)^{-1}$, $\omega_S$ is the probe frequency and $\Gamma$ depends on $\omega_S$ and the network Hamiltonian $H_E$. When $T=0$ the latter term on the R.H.S. of Eq.~\eqref{eq:excitations} vanishes and consequently $\Delta\langle n\rangle=|\langle a^\dagger(t)a(t)\rangle-\langle a^\dagger(0)a(0)\rangle|$ becomes directly proportional to $\langle a^\dagger(0)a(0)\rangle$. Then changing probe initial excitations can only scale $\Delta\langle n\rangle$ and therefore can be accounted for by scaling the values of $\Delta\langle n\rangle$ in some fixed way before normal mode frequencies are extracted as described in \ref{app:quantum}. Assuming Markovian dynamics and that the finite size effects are not too strong the estimate is then very robust to changes in initial conditions.

To verify that the analytic expression holds to a good enough approximation in spite of the complex normal mode structure (which in general can lead to non-Markovian dynamics) and the finite size of the network, we considered several different initial states for the probe. The network is the same as in Fig.~\ref{fig:5}, and the spectral dimension is estimated following exactly the same steps, except that $\Delta\langle n\rangle$ is scaled such that its maximum value is $35$, which is approximately the same as in the left panel of Fig.~\ref{fig:5}. This time we consider only the case where the probe is coupled to 8 randomly chosen network nodes and 10 frequency sweeps are made.

\begin{figure}[t]
\centering
                \includegraphics[trim=0cm 0cm 0cm 0cm,clip=true,width=0.95\textwidth]{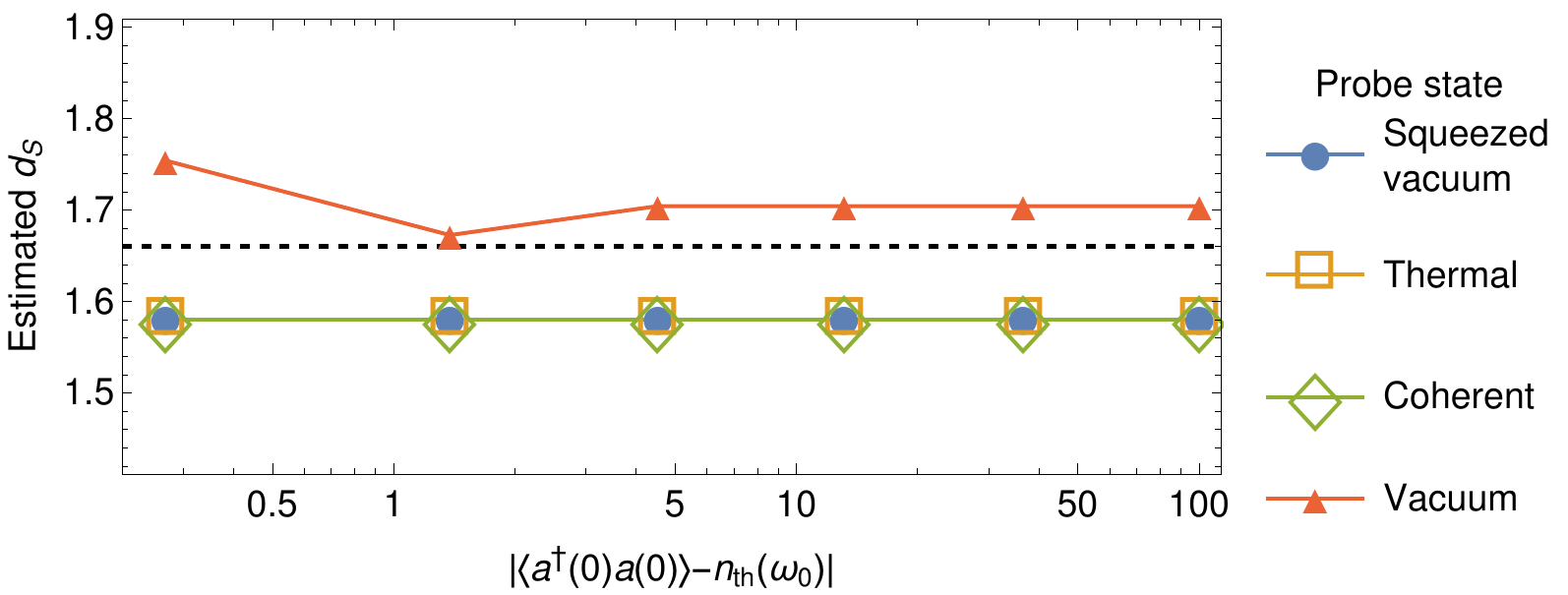}
            \caption{\label{fig:6}
Impact of different initial conditions on the estimated spectral dimension. When the probe initial state is the vacuum the network temperature $T>0$, while otherwise $T=0$. The value of $d_S$ estimated directly from the full set of normal mode frequencies is shown by the horizontal dashed line.
}
      \end{figure}

The results are shown in Fig.~\ref{fig:6} where the estimated spectral dimension $d_S$ is compared to the difference between probe initial excitations and $n_{th}(\omega_0)$, where $\omega_0$ is the network bare frequency. Focusing first on the cases where the probe initial state is not the vacuum ($T=0$), we can see that neither the state itself nor the initial excitations affect the estimate, as expected. We also considered the case where the probe is initially in vacuum and $T>0$. In this case there is some variation in the estimate for small and intermediate temperatures because $n_{th}(\omega_S)$ in Eq.~\eqref{eq:excitations} depends on probe frequency, but at high temperatures this effect becomes negligible.

For other network states, such as multi-mode squeezed states, the last term of the R.H.S. of Eq.~\eqref{eq:excitations} is replaced by a more general term that depends in particular of the probe frequency \cite{maniscalco2004simulating}. Provided that this dependency is not significantly stronger than in the case of a thermal state our results remain directly applicable.  

\section*{References}

\bibliography{references}

\bibliographystyle{unsrt-abbrv}

\end{document}